\documentclass[fleqn,usenatbib]{aa}

\usepackage{graphicx}
\usepackage{gensymb}
\usepackage{multirow}
\usepackage{subcaption}
\usepackage{txfonts}
\usepackage{lineno}
\usepackage{colortbl}
\usepackage[normalem]{ulem}
\usepackage{natbib}
\bibpunct{(}{)}{;}{a}{}{,}
\usepackage[
    bookmarks,
    bookmarksopen=true,
    bookmarksnumbered=true,
    breaklinks=true,
    colorlinks=true,
    linkcolor=blue,
    urlcolor=blue,
    citecolor=blue,
    anchorcolor=blue,
    hyperindex=true,
    hyperfigures
]{hyperref}

\usepackage{xcolor}
\definecolor{footnotelinkcolor}{RGB}{200, 0, 0}  

\usepackage[table,xcdraw]{xcolor}
\begin{document} 
   \title{CAPOS: The bulge Cluster APOGEE Survey VI. }

   \subtitle{Characterizing multiple stellar populations and chemical abundances in the bulge globular cluster NGC 6569.}

   \author{Nicolás Barrera
          \inst{1,2}\fnmsep\thanks{e-mail: nicolas.barrerad@userena.cl}
          \and
          Sandro Villanova\inst{3}
          \and
          Doug Geisler \inst{1,2}
          \and
          José G. Fernández-Trincado \inst{4}\fnmsep\thanks{e-mail: jose.fernandez@ucn.cl}
          \and
          Cesar Muñoz \inst{2}
          }

   \institute{
        Departamento de Astronomía, Casilla 160-C, Universidad de Concepción, Concepción, Chile
         \and 
         Departamento de Astronomía, Facultad de Ciencias, Universidad de La Serena, Av. Juan Cisternas 1200, La Serena, Chile
         \and 
         {Universidad Andres Bello, Facultad de Ciencias Exactas, Departamento de F{\'i}sica y Astronom{\'i}a - Instituto de Astrof{\'i}sica, Autopista Concepci\'on-Talcahuano 7100, Talcahuano, Chile}
         \and 
         Instituto de Astronomía, Universidad Católica del Norte, Av. Angamos 0610, Antofagasta, Chile
         }

   \date{Received XXXX; accepted XXXX}

\abstract
{The CAPOS project aims to obtain accurate mean abundances for many elements and their mean radial velocities, and it explores the multiple population (MP) phenomenon in Galactic bulge globular clusters (BGCs). NGC 6569 is one of the clusters observed by CAPOS.}
{This study presents a detailed high-resolution spectroscopic analysis of NGC 6569 to derive high-precision mean abundances for a number of elements with various nucleosynthetic origins and to unveil its MPs by focusing on key spectral features. Our aim is to complement previous suggestions of the presence of MPs in this cluster based on the typical Na–O anticorrelation and the presence of a double horizontal branch.}
{We analyzed the near-infrared APOGEE-2 spectra of 11 giant member stars in NGC 6569 using the code BACCHUS. We derived abundances for 12 elements, including light elements (C, N), $\alpha$-elements (O, Mg, Si, Ca, Ti), iron-peak elements (Fe, Ni), the odd-Z element (Al), and s-process elements (Ce, Nd). We also performed an isochrone fitting using photometric data (Gaia + 2MASS) to estimate atmospheric parameters, the cluster distance, and its extinction.}
{We derived a mean metallicity of [Fe/H] = –0.91 ± 0.06, which is consistent with the values from the APOGEE pipeline and slightly more metal poor than previous findings. The scatter lies within the observational uncertainties. 
The cluster shows enhanced $\alpha$-element abundances ($[\alpha/\text{Fe}] = 0.36 \pm 0.06$ dex) similar to other Galactic globular clusters (GCs). We find no significant variation in Al, suggesting a homogeneous distribution within the cluster. In contrast, we find considerable N-enrichment ([N/Fe] = 0.68 ± 0.34 dex) and a large spread of 0.90 dex, which enabled us to distinguish at least two separate populations based on N that have anticorrelated C abundances. The n-capture elements Ce and Nd are overabundant compared to the Sun, but are similar to those of GCs in this metallicity regime, and also show an average ratio of $\langle [\text{Ce/Nd}] \rangle = -0.17 \pm 0.12$. Finally, we estimated a mean radial velocity of RV = –49.75 ± 3.68 km s$^{-1}$, which is consistent with previous measurements, but the heliocentric distance ($\text{d}_\odot$ = 12.4 ± 1.45 kpc) and interstellar reddening (E(B–V) = 0.68) are higher than reported in the literature.}
{The analysis confirms the presence of MPs in NGC 6569, evidenced by a significant spread in N and a clear C–N anticorrelation. This supports the previously established Na–O anticorrelation. MPs are characterized through this pattern for the first time. NGC 6569 exhibits chemical signatures typical of BGCs, without a significant spread in metallicity. The cluster $\alpha$-element enhancement (consistent with early enrichment by type II supernovae) and the absence of a Mg–Al–Si anticorrelation agree with expectations for relatively high-metallicity GCs and suggests a rapid and homogeneous star formation history. The overabundance of n-capture elements indicates contributions from r-process events and might be linked to neutron star mergers. These n-capture elements are reported in NGC 6569 for the first time.}

   \maketitle

\section{Introduction}

Nearly all Galactic globular clusters (GCs) have been found to host multiple stellar populations (MPs) that are identified by star-to-star variations in the light (C, N), odd-Z (Na, Al), $\alpha$ (O, Mg), and neutron-capture (Ce) elements \citep[e.g., ][]{gratton2004abundance,2008ApJ...672L..29Y,Yong_2009,2009A&A...505.1099M,2009A&A...505..117C,carretta2009anticorrelation,Carretta_2010,Villanova_2010,pancino2017gaia,2019A&A...622A.191M,2020MNRAS.492.1641M,2021A&A...652A.157G,2021A&A...652A.158R,2022A&A...658A.116F,ian2022,2023MNRAS.526.6274G}. This revelation has challenged our understanding of the GC formation and evolution for decades because it suggests a complex chemical enrichment history that is not expected from standard stellar evolution, that is, simple stellar populations \citep{gratton2012,2017A&A...607A..44B,Bastian_2018}. They also offer a unique window into the early evolutionary processes of GCs, however, and provide clues about the conditions in the early Universe that shaped these stellar systems \citep{Renzini2015}. The general scenario suggests that an initial generation of stars formed from gas that was previously enriched by type II supernovae, which contributes to an overabundance of $\alpha$ elements and to a relative deficiency in other light elements such as Na and Al \citep{Carretta_2018}. This initial generation of stars is often referred to as the first population (1P), and it is characterized by a chemical composition resembling that of field stars with a similar metallicity. A second population (2P) of stars forms from a mix of self-enriched stellar ejecta and pristine gas, which exhibits well-known anticorrelations such as C–N, Na–O, and Mg–Al \citep[e.g., ][]{1987PASP...99...67S,carretta2009anticorrelation,2015AJ....149..153M}. The near ubiquity of these chemical signatures clearly suggests that this sequence of events is essential to the formation of GCs, but the exact nature of the first-generation polluters (e.g., intermediate-mass asymptotic giant branch (AGB) stars; \citep{1981ApJ...245L..79C, 2001ApJ...550L..65V}; fast-rotating massive stars; \citet{2007A&A...464.1029D}; massive binaries; \citet{2009A&A...507L...1D}; supermassive stars; \citet{Gieles_2019}) and the mechanisms driving the secondary star formation remain unknown \citep[see][for an extended discussion]{Bastian_2018}.

The presence of MPs has been identified in disk and bulge globular clusters \citep[BGCs; see, e.g.,][]{Carretta_2018,2019A&A...627A.178F, 2022A&A...658A.116F,2023MNRAS.526.6274G}, which are thought to have developed in situ \citep{2019A&A...630L...4M}, and in their halo counterparts \citep[e.g., ][]{gratton2004abundance,2009A&A...505..117C,2015AJ....149..153M,ian2022}, most of which originated in satellite galaxies that later merged with the proto-Milky Way \citep[see, e.g., ][]{Forbes_2010,Kruijssen_2018,2018ApJ...863L..28M}. However, our knowledge of the former has remained mostly hidden by high interstellar extinction and stellar crowding toward these regions, where only a limited number of BGCs with relatively high metallicity have been investigated in detail \citep[see, e.g., ][]{Nataf_2019}. Large-scale spectroscopic surveys such as the Apache Point Observatory Galactic Evolution Experiment \citep[APOGEE, ][]{2017AJ....154...94M} of the Sloan Digital Sky Survey-IV \citep[SDSS-IV, ][]{2017AJ....154...28B} have helped us to improve the limited sample of BGCs and the identification of MPs because the APOGEE near-infrared (NIR) spectral range significantly minimizes the effects of dust obscuration. In this context, along with the high-precision astrometric and photometric data from the latest release of the ESA Gaia mission \citep[Gaia DR3, ][]{2023A&A...674A...1G}, combined with the photometric depth of the Two Micron All Sky Survey \citep[2MASS, ][]{2006AJ....131.1163S} and the infrared survey VISTA Variables in the Vía Láctea \citep[VVV, ][]{MINNITI2010433,2012A&A...537A.107S}, the main goal of the bulge Cluster APOgee Survey \citep[CAPOS, ][]{2021A&A...652A.157G} is to considerably augment SDSS-IV BGC studies and help us to explore the MP phenomenon at the high-metallicity end. CAPOS uses the capabilities of the 16th and 17th APOGEE-2 data releases \citep{2020ApJS..249....3A,2022ApJS..259...35A} and observed a total of 18 BGCs. An overview and initial results for the first sample of BGCs observed by CAPOS were given in \citet{2021A&A...652A.157G}. A second study \citep{2021A&A...652A.158R} presented results from CAPOS observations of the recently discovered intriguing BGC FSR-1758. The third paper in the series was the first high-resolution study of Ton 2 \citep{2022A&A...658A.116F}, and CAPOS IV \citep{2023MNRAS.526.6274G} explored the chemical composition of NGC 6558. More recently, the fifth CAPOS paper presented the detailed chemical abundances of the BGC HP1 \citep{2025A&A...696A.154H}. This paper is the sixth in the CAPOS series.

The object of this study, NGC 6569, is another BGC observed by CAPOS. NGC 6569 is an old \citep[12.8 $\pm$ 0.1 Gyr, ][]{2019ApJ...874...86S}, massive \citep[$2.29 \pm 0.21 
\times10^5 \text{M}_\odot$, ][]{2021MNRAS.505.5957B} globular cluster located in the Galactic bulge ($l^{\circ}, b^{\circ} = 0.487, -6.676$) at approximately 3 kpc from the Galactic center \citep{2021MNRAS.505.5957B} and $\approx 10$ kpc from the Sun \citep{2019ApJ...874...86S}. It has a relatively narrow range of metallicity estimates, [Fe/H] $\sim$ –0.87 to –0.76 \citep{1985ApJ...293..424Z,bica1983ddo,2005MNRAS.361..272V,2011MNRAS.414.2690V,2018AJ....155...71J} that come from deep photometry and high-resolution spectroscopy in a region with high foreground interstellar reddening, E(B–V) $\sim$ 0.49–0.55 \citep{bica1983ddo,1985ApJ...293..424Z,2005MNRAS.361..272V}.

Recently, \citet{2023AAS...24140217H} have identified at least 23 extratidal stars at up to 5 tidal radii from the cluster, with radial velocities and metallicities that match those of NGC 6569, suggesting significant tidal interactions with the Galactic bulge and bar. The cluster orbit indicates significant influence from the Galactic bar, as reflected by a capture probability of 0.2 \citep{2023AstBu..78..499B}, which is notably high compared to typical values in the inner bulge, which are usually below 0.1. The integration of its orbit in a nonaxisymmetric potential shows a regular stable motion over long periods that supports the impact of the bar.

Photometrically, this cluster exhibits a peculiar well-studied double horizontal branch \citep{2012ApJ...761L..29M}, with two distinct groups of stars that are separated by about 0.1 magnitudes in the $K_s$ band, suggesting the presence of MPs within the cluster. The difference in luminosity is likely due to slight variations in helium content rather than differences in metallicity or other light-element abundances, indicating intrinsic features of the cluster rather than observational errors or field contamination \citep{2018AJ....155...71J}.

On the other hand, high-resolution near-infrared spectroscopy has revealed the chemical composition of NGC 6569, and a relatively high metallicity ([Fe/H] = –0.79 and –0.87) and enhanced $\alpha$-elements ([$\alpha$/Fe] = 0.43 and 0.34) were estimated by \citet{2011MNRAS.414.2690V} and \citet{2018AJ....155...71J}, respectively. Moreover, a clear Na–O anticorrelation and low $^{12}$C/$^{13}$C ratios suggest at least two distinct stellar populations, indicating additional mixing during the red giant branch phase and supporting the evidence of MPs in NGC 6569 \citep{2018AJ....155...71J}. Additionally, \citet{2018AJ....155...71J} found that NGC 6569 exhibits significant enrichment in heavy elements, with [La/Fe] = +0.38 dex and [Eu/Fe] = +0.49 dex. The moderately depleted mean [La/Eu] ratio of –0.11 dex suggests significant pollution via the r-process, although the cluster also experienced some s-process enrichment, as indicated by the higher [La/Eu] values compared to more metal-poor systems \citep[e.g., ][]{Yong_2005}. We present a high-resolution spectroscopy study of 11 members of NGC 6569 from APOGEE-2 spectra as part of CAPOS in order to further investigate the detailed abundances and MPs.

This paper is organized as follows: We detail in Section \ref{data} the observational data we used. In Section \ref{cluster}, we present the cluster membership analysis. Section \ref{atmosparam} describes the different atmospheric parameters under consideration, and Section \ref{elementalabundances} presents the resulting elemental abundances. Finally, we summarize our work in Section \ref{conclusion}.

\section{Data and sample}\label{data}

This study is based on high-resolution (R $\sim$ 22,500) NIR spectra obtained by the Apache Point Observatory Galactic Evolution Experiment II survey \citep[APOGEE-2, ][]{2017AJ....154...94M}, an internal program of SDSS-IV \citep[]{2017AJ....154...28B}. APOGEE-2 is designed to provide precise radial velocities (RV < 1 km s$^{-1}$) and detailed chemical abundances for a large sample of giant stars, with the goal of unveiling the dynamical structure and chemical history of the Milky Way galaxy.

APOGEE-2 spectra are observed with custom-built multi-object spectrographs \citep{2019PASP..131e5001W}, located in the Northern Hemisphere on the 2.5 m telescope at Apache Point Observatory (APO; APOGEE-2N; \citealt{2006AJ....131.2332G}) and in the Southern Hemisphere on the Irénée du Pont 2.5 m telescope \citep{1973ApOpt..12.1430B} at Las Campanas Observatory (LCO; APOGEE-2S), capable of simultaneously recording the spectra of up to 300 targets. Each instrument records most of the H band (1.51 $\mu$m – 1.69 $\mu$m) on three detectors, with coverage gaps between $\sim$1.58–1.59 $\mu$m and $\sim$1.64–1.65 $\mu$m. Each fiber subtends an on-sky field of view with a diameter of $\sim$ 2''  in the northern and 1.3'' in the southern instrument.

The final release of APOGEE-2 data from SDSS-III/SDSS-IV, DR17, includes all data collected at APO through November 2020 and at LCO through January 2021. The dual APOGEE-2 instruments observed over 650,000 stars across the Milky Way. The target selection was detailed in \citet{2017AJ....154..198Z}, \citet{2021AJ....162..302B}, and \citet{2021AJ....162..303S}. The spectra were reduced as described by \citet{2015AJ....150..173N} and were analyzed using the APOGEE Stellar Parameters and Chemical Abundance Pipeline \citep[ASPCAP, ][]{GarcíaPérez_2016} and the libraries of synthetic spectra described by \citet{2015AJ....149..181Z}. The customized H-band line lists were fully described in \citet{2015ApJS..221...24S}, \citet{2016ApJ...833...81H} neodymium lines (Nd II), \citet{2017ApJ...844..145C} cerium lines (Ce II), and \citet{2021AJ....161..254S}.

\begin{figure*}[htbp!]
    \centering
    \subfloat{
        \includegraphics[width=0.45\textwidth, height=0.3\textheight]{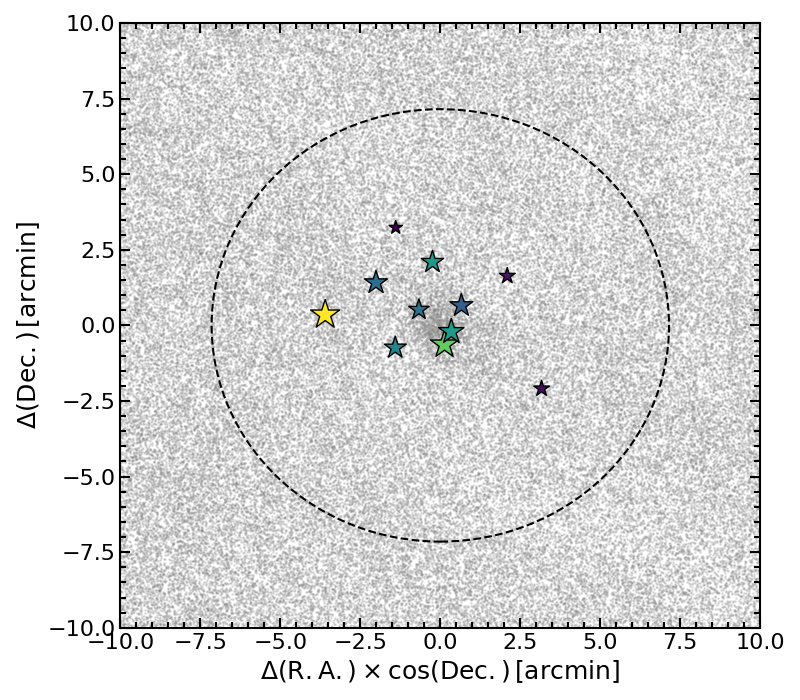}
    }
    \subfloat{
        \includegraphics[width=0.45\textwidth, height=0.3\textheight]{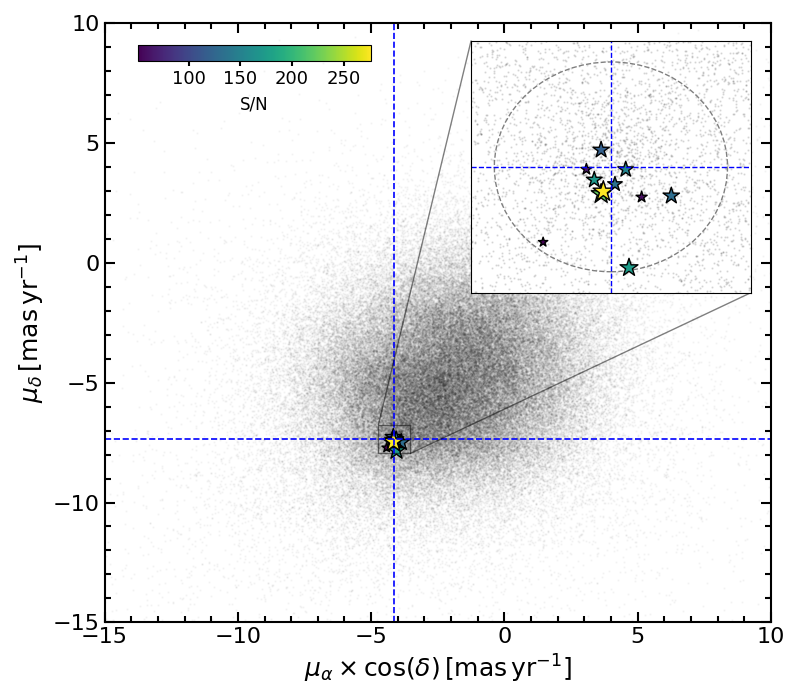}
    }\\
    \subfloat{
        \includegraphics[width=0.465\textwidth, height=0.3\textheight]{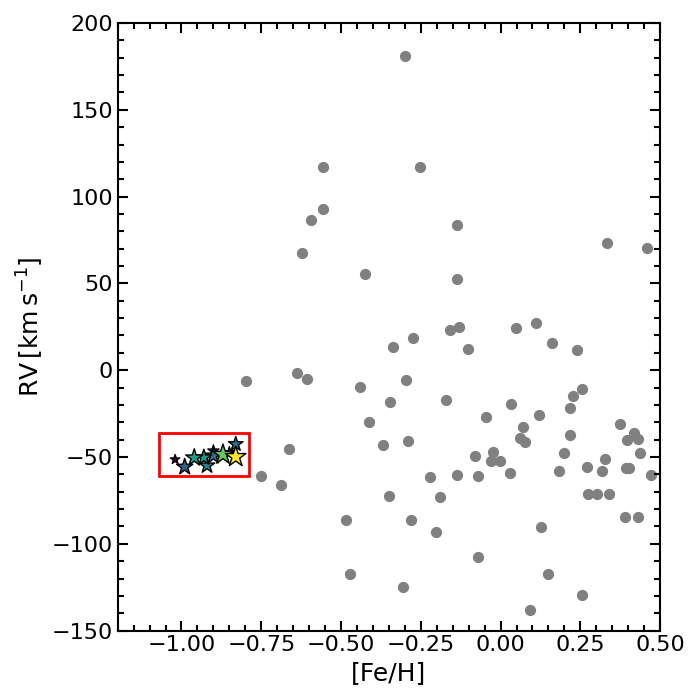}
    }
    \subfloat{
        \includegraphics[width=0.44\textwidth, height=0.3\textheight]{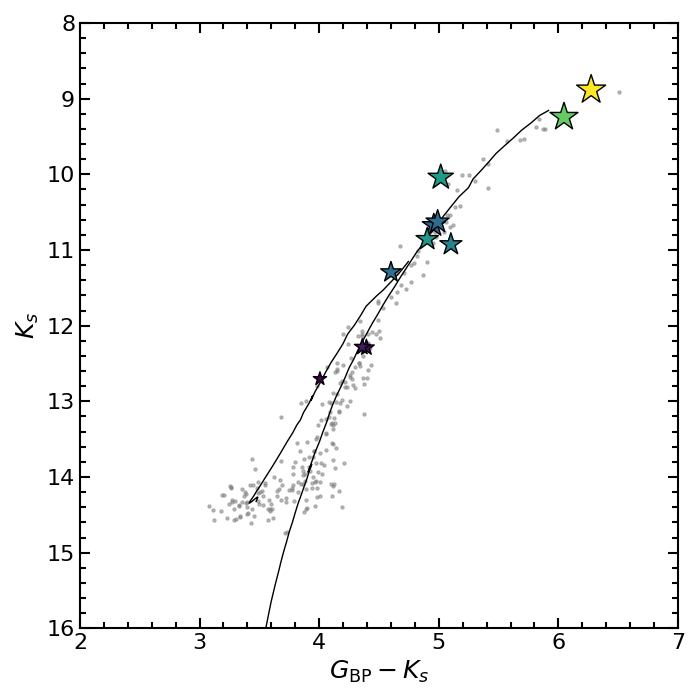}
    }
    \caption{Global properties of NGC 6569 targets. \textit{Top left} panel: Spatial position. The color-coded symbols represent the S/N of stars (as labeled in the \textit{top right} panel) with APOGEE spectra, whose sizes are proportional to their K$_s$ magnitudes. Field stars are shown as gray dots. A circle with the tidal radius of 7.15' \citep{2010arXiv1012.3224H} is overplotted. \textit{Top right} panel: Proper motion density distribution of stars located within the tidal radius from the cluster center, color- and size-coded as in the \textit{top left} panel. The inset in the top right panel shows a zoom-in of the cluster, squared in 0.7 × 0.7 mas yr$^{-1}$ and enclosed in a proper motion radius of 0.5 mas yr$^{-1}$, shown as a gray circle. The dashed blue lines are centered on the nominal proper motion values from \citet{2021MNRAS.505.5957B}. \textit{Bottom left} panel: Radial velocity vs. metallicity of our members compared to field stars. The [Fe/H] values for our targets have been determined with BACCHUS (see Section \ref{elementalabundances}) using photometric atmospheric parameters (see Table \ref{parameters}), while the [Fe/H] values for field stars are from the ASPCAP pipeline (gray dots). The red box encloses the cluster members within 0.11 dex and 9 km s$^{-1}$ from the nominal mean [Fe/H] = –0.91 dex and RV = –49.75 km s$^{-1}$ of NGC 6569, as determined in this work. \textit{Bottom right} panel: Color–magnitude diagram corrected for differential reddening and extinction in the Gaia BP band and 2MASS K band of our sample and stars within 7.15'. The best PARSEC isochrone fit of 12 Gyr is shown as a solid black line for NGC 6569 stars (gray dots) located within 7.15' from the cluster center.
}
\label{starsdiagram}
\end{figure*}

\section{NGC 6569: Membership selection}\label{cluster}
NGC 6569 was observed as part of the contributed APOGEE-2S CNTAC\footnote{Chilean Telescope Allocation Committee} CN2019A--98 program (P.I.: Geisler) during July 9–10, 2019, as part of the CAPOS survey \citep{2021A&A...652A.157G}. The APOGEE-2S plug-plate containing NGC 6569 was centered on ($l$, $b$) $\sim$ (0.48\degree, –6.68\degree), inside a radius of $\leq$ 15' from the cluster center. An additional GC, NGC 6558, was observed simultaneously and will be reported elsewhere. In order to maximize membership probability, we restricted our selection of final potential members to targets inside the tidal radius of 7.15' \citep{2010arXiv1012.3224H}, as shown in Figure~\ref{starsdiagram}. We decided to exclude possible extratidal stars, even though several are identified in NGC 6569 \citep{2023AAS...24140217H}, because their position in the CMD makes them unlikely members.

The CAPOS targets to observe were originally selected prior to Gaia. Post-observations, an additional astrometric membership criterion was applied based on Gaia DR2 \citep{2018A&A...616A...1G} and required proper motions (PMs) within a radius of 0.5 mas yr$^{-1}$ around the nominal PM values of NGC 6569: $\mu_\alpha \cos(\delta)$ = –4.125 mas yr$^{-1}$ and $\mu_\delta$ = –7.354 mas yr$^{-1}$ \citep{2021MNRAS.505.5957B}. We initially used these values to estimate candidate members from the PMs and to reduce potential field star contamination in the optical and NIR color–magnitude diagrams (CMDs), but we reexamined the PMs from the original sample using the values in Gaia DR3 \citep{2023A&A...674A...1G} toward NGC 6569, as shown in Fig.~\ref{starsdiagram}. The original astrometric members remained, and no additional members were found.

Our targets are positioned from near the tip of the red giant branch (RGB), as shown in the CMD corrected for differential reddening (see Fig.~\ref{starsdiagram}), to the 2MASS K$_\text{s}$ band brighter than 12.7 mag. This was required in order to achieve a minimum signal-to-noise ratio of S/N > 60 pixel$^{-1}$ in one plug-plate visit ($\sim$1 hour). Although more visits were originally planned, because of weather, time allocation, and airmass constraints, only one visit was obtained in the end. Ten out of the eleven observed stars reached S/N > 60 pixel$^{-1}$, and the remaining spectrum has a lower S/N of 51 pixel$^{-1}$, which is the hottest star (4526.9 K) (see Table~\ref{ngc6569membersastrometric}).

In the following, we use all stars to provide reliable and precise (< 1 km s$^{-1}$) radial velocities for the cluster membership confirmation, and although it does not satisfy the minimum S/N threshold used in previous CAPOS clusters \citep{2021A&A...652A.158R, 2022A&A...658A.116F,2023MNRAS.526.6274G}, we decided to include our low S/N star for the abundance analysis because we were able to obtain sufficiently reliable measurements ($\sigma < 0.1$; see Eq.~\ref{error}). The CAPOS minimum S/N ratio has now been lowered to 50, so that our lowest S/N star satisfies this criterion. These stars are highlighted in Figure~\ref{starsdiagram}.

Figure~\ref{starsdiagram} shows the BACCHUS [Fe/H] abundance ratios (see Section~\ref{elementalabundances}) versus the radial velocity of our 11 potential cluster members compared to field stars with ASPCAP/APOGEE-2 [Fe/H] determinations, which were shifted by +0.11 dex in order to minimize the systematic differences between ASPCAP/APOGEE-2 and BACCHUS, as highlighted in Appendix D of \citet{2020A&A...643L...4F}. We note that all of our targets have very similar radial velocities, which indicates a strong likelihood of cluster membership. This identification is further reinforced by the fact that their metallicities are tightly clustered and clearly distinct from the surrounding field population. 

We find a mean RV from 11 APOGEE-2 stars of –49.75 $\pm$ 3.68 km s$^{-1}$, which agrees excellently with the value listed in Baumgardt’s catalogue\footnote{\hypersetup{urlcolor=footnotelinkcolor}\url{https://people.smp.uq.edu.au/HolgerBaumgardt/globular/}\hypersetup{urlcolor=blue}}
, RV = –49.83 $\pm$ 0.50 km s$^{-1}$, and \citet{2018AJ....155...71J}, RV = –48.80 $\pm$ 5.30 km s$^{-1}$. Other sources inside this box are nonmembers (red circles) with properties that are not compatible with the cluster. The red box highlighted in Figure~\ref{starsdiagram} encloses the cluster members within 0.11 dex and 9 km s$^{-1}$ from the nominal mean [Fe/H] = –0.91 dex and RV = –49.75 km s$^{-1}$ of NGC 6569, as determined in this work (see Section~\ref{elementalabundances}).

\begin{table*}[ht!]
    \caption{Astrometric, photometric, and kinematic properties of NGC 6569 members.}
    \centering
    \renewcommand{\arraystretch}{1.25}
    \scalebox{0.8}{
    \begin{tabular}{l|l|l|l|l|l|l|l|l}
        \hline \hline
        APOGEE ID & $\alpha$ & $\delta$ & $\text{G}_{BP}$ & $\text{K}_s$ & $\mu_\alpha\cos\delta$ & $\mu_\delta$ & RV & S/N \\ 
         & [deg] & [deg] & [mag] & [mag] & [mas/yr] & [mas/yr] & [km\,s$^{-1}$] & [pixel$^{-1}$] \\ \hline \hline
2M18132448-3149140 & 273.352 & -31.821 & 15.16 & 8.88 & -4.158 & -7.472 & -49.94 & 276 \\ 
2M18133940-3150132 & 273.414 & -31.837 & 15.29 & 9.24 & -4.167 & -7.483 & -48.64 & 222 \\ 
2M18134025-3149477 & 273.418 & -31.830 & 15.24 & 10.36 & -4.048 & -7.835 & -50.45 & 174 \\ 
2M18133789-3147295 & 273.408 & -31.792 & 15.76 & 10.85 & -4.196 & -7.416 & -50.32 & 171 \\ 
2M18133324-3150194 & 273.389 & -31.839 & 16.03 & 10.90 & -4.062 & -7.366 & -54.96 & 153 \\ 
2M18133620-3149040 & 273.401 & -31.818 & 15.91 & 11.30 & -4.108 & -7.436 & -42.58 & 136 \\ 
2M18133083-3148103 & 273.378 & -31.803 & 15.67 & 10.63 & -3.867 & -7.492 & -49.30 & 130 \\ 
2M18134151-3148556 & 273.423 & -31.815 & 15.67 & 10.68 & -4.167 & -7.273 & -55.72 & 119 \\ 
2M18134725-3147570 & 273.447 & -31.799 & 16.64 & 12.28 & -4.230 & -7.366 & -46.46 & 64 \\ 
2M18135154-3151406 & 273.465 & -31.861 & 16.70 & 12.29 & -3.993 & -7.498 & -47.48 & 60 \\ 
2M18133329-3146211 & 273.389 & -31.773 & 16.76 & 12.71 & -4.416 & -7.713 & -51.42 & 51 \\ \hline \hline
    \end{tabular}}
    \label{ngc6569membersastrometric}
\end{table*}

The CMD presented in Figure~\ref{starsdiagram} was corrected for differential reddening using giant stars, following the same method as employed by \citet{2022A&A...658A.116F}. For this purpose, we selected all RGB stars within a radius of 3.5 arcminutes from the cluster center whose proper motions were compatible with that of NGC 6569. First, we drew a ridge line along the RGB, and we calculated the distance from this line for each of the selected RGB stars along the reddening vector. The vertical projection of this distance gives the differential interstellar absorption at the position of the star, and the horizontal projection gives the differential optical+NIR reddening at the position of the star. 

After this first step, we selected the three nearest RGB stars for each star in the field, calculated the mean interstellar reddening and absorption, and finally, subtracted these mean values from its optical+NIR colors and magnitudes. 

We underline the fact that the number of reference stars used for the reddening correction is a compromise between a correction that is affected as little as possible by photometric random error and the highest possible spatial resolution.

In order to estimate the distance of the cluster, we performed an isochrone fitting of the RGB using the PARSEC database\footnote{\hypersetup{urlcolor=footnotelinkcolor}\url{http://stev.oapd.inaf.it/cgi-bin/cmd}\hypersetup{urlcolor=blue}}
 \citep{2012MNRAS.427..127B}. The extinction correction was applied to the isochrone using the extinction laws of \citet{1989ApJ...345..245C} and \citet{1994ApJ...422..158O}. 

The free parameters for this fitting are the true distance modulus, (m–M)$_0$, the interstellar absorption in the V band, A$_\text{V}$, the reddening-law coefficient, R$_\text{V}$, and the global metallicity [M/H]. These four parameters were estimated simultaneously by fitting the B$_\text{BP}$–K$_s$ versus K$_s$, B$_\text{BP}$–R$_\text{BP}$ versus B$_\text{BP}$, and J$_s$–K$_s$ versus K$_s$ CMDs. Figure~\ref{starsdiagram} shows the best-fitting isochrone on the G$_\text{RP}$–K$_s$ versus K$_s$ CMD. For this fitting, an age of 12 Gyr was assumed, and a global metallicity of [M/H] = –0.80 was found. This global metallicity is higher than the [Fe/H] found in Section~\ref{elementalabundances} because it includes both the [Fe/H] content and the $\alpha$-enhancement of the cluster.

The extinction-law coefficient R$_\text{V}$ is usually assumed to be 3.1 in the solar neighborhood, but it can vary significantly from this canonical value, especially in the direction of the Galactic bulge \citep{2016MNRAS.456.2692N}, where it can easily take lower values. We found an extinction-law coefficient of R$_\text{V}$ = 2.2. Moreover, we achieved a fit for the distance $\text{d}_\odot$ = 12.4 $\pm$ 1.45 kpc, whic is somewhat higher than (but still compatible within the errors with) previous estimations: 10.53 kpc \citep{2021MNRAS.505.5957B}, 10.1 kpc \citep{2019ApJ...874...86S}, and 10.9 kpc \citep{2010arXiv1012.3224H}; and an interstellar absorption of A$_\text{V}$ = 1.5. 

Finally, the interstellar absorption and extinction-law coefficient we found can be translated into E(B–V) = 0.682 for NGC 6569, which is higher than the foreground interstellar reddening determined by previous works: E(B–V) $\sim$ 0.49–0.55 \citep{bica1983ddo,1985ApJ...293..424Z,2005MNRAS.361..272V,2010arXiv1012.3224H}. The discrepancies in the distance modulus and E(B–V) reddening are due to the lower extinction-law coefficient.

\section{Stellar atmospheric parameters and chemical abundance measurements}\label{atmosparam}
In order to derive elemental abundances, we inspected each APOGEE spectrum. We made use of the code called Brussels Automatic Stellar Parameter \citep[BACCHUS; ][]{2016ascl.soft05004M} to derive the metallicity, broadening parameters, and chemical abundances for the entire sample, making a careful line selection as well as providing abundances based on a line-by-line differential approach. 

The BACCHUS module consists of a shell script that computes synthetic spectra on the fly for a range of abundances and compares these syntheses to observational data on a line-by-line basis, deriving abundances from different methods (e.g., using the equivalent width, the line depth, or the $\chi^2$). For the synthetic spectra, BACCHUS relies on the radiative transfer code Turbospectrum \citep{1998A&A...330.1109A,2012ascl.soft05004P} and the MARCS model atmosphere grid \citep{2008A&A...486..951G}. 

From these synthetic spectra, the BACCHUS code then identifies the continuum (or pseudo-continuum) points for normalizing the observed spectra and the relevant pixels (i.e., a mask or window) to use for the abundance determination of each line in each spectrum. The normalization of the observed spectra is performed by selecting continuum wavelengths in the synthetic spectra and fitting a linear relation across these continuum points within 30~\r{A} around the line of interest in the observed spectra. 

The spectral window for each line is determined by comparing where changes in elemental abundance produce significant changes in the synthetic spectra with the second derivative of the observed flux (i.e., to identify the local maxima on either side of the line of interest).

For each element and each line, the abundance determination proceeds as described by \citet{2016A&A...594A..43H,2017ApJ...846L...2F,2019MNRAS.488.2864F,2019A&A...631A..97F,2019ApJ...886L...8F,2020A&A...643L...4F,2020A&A...644A..83F,2020ApJ...903L..17F,2021ApJ...918L...9F,2021A&A...652A.158R}, and we therefore only provide a brief overview here. The abundance of each chemical species was computed as follows: (a) A synthesis was performed using the full set of atomic and molecular line lists described by \citet{2015ApJS..221...24S}, \citet{2016ApJ...833...81H}, \citet{2017ApJ...844..145C}, and \citet{2021AJ....161..254S}. This set of lists was internally labeled as \textit{linelist.20170418}, based on the date of creation in the format YYYYMMDD. This was used to find the local continuum level via a linear fit. (b) We rejected cosmic ray and telluric lines. (c) The local S/N was estimated. (d) A series of flux points contributing to a given absorption line were automatically selected. Finally (e) the abundances were derived by comparing the observed spectrum with a set of convolved synthetic spectra characterized by different abundances. Then, four different abundance determination methods were used as listed below.

\begin{enumerate}
    \item Line-profile fitting: This method, also known as the \(\chi^2\) method, determines an abundance by minimizing the squared differences between synthetic and observed spectra. It seeks the best fit to the entire line profile and provides a comprehensive assessment of the spectral match.
    
    \item Core line-intensity comparison: This method, referred to as the int method, focuses on matching the intensity at the core of the line between the synthetic and observed spectra. It is particularly useful for analyzing the central features of spectral lines where the abundance signature is most prominent.
    
    \item Global goodness-of-fit estimate: Known as the syn method, this approach determines the abundance that minimizes the overall difference between the synthetic and observed spectra by evaluating the global quality of the spectral fit instead of focusing on individual line features.
    
    \item Equivalent-width comparison: This method, called the eqw method, determines the abundance required to match the equivalent width of the synthetic spectrum with that of the observed spectrum. It is a commonly used technique that simplifies the comparison by reducing the spectral information to a single value representing the area below the line profile.

\end{enumerate}
Each diagnostic yields validation flags. Based on these flags, a decision tree then rejects or accepts each estimate, keeping the best-fit abundance. We adopted the $\chi^2$ diagnostic as the final abundance because it is the most robust \citep{2016A&A...594A..43H}. However, we stored the information from the other diagnostics, including the standard deviation among all four methods.

The CNO elements require special treatment in this abundance analysis. The abundances of $^{12}$C, $^{14}$N, and $^{16}$O were derived through a combination of vibration–rotation lines of $^{12}$C$^{16}$O and $^{16}$OH, along with electronic transitions of $^{12}$C$^{14}$N. The procedure for the CNO analysis begins by setting the C abundance using $^{12}$C$^{16}$O lines. With this C abundance, the O abundance is determined using $^{16}$OH lines. When this O abundance differs from the initial value (which is scaled from the solar value by the stellar [Fe/H] ratio), the $^{12}$C$^{16}$O lines are reanalyzed with the updated O abundance. 

This iterative process continues until consistent C and O abundances are obtained from the $^{12}$C$^{16}$O and $^{16}$OH lines. When consistent values for C and O are established, the $^{12}$C$^{14}$N lines are used to derive the N abundance. Generally, the N abundance has little to no effect on the $^{12}$C$^{16}$O and $^{16}$OH lines, but the final C, N, and O abundances provide self-consistent results from $^{12}$C$^{16}$O, $^{16}$OH, and $^{12}$C$^{14}$N \citep{2013ApJ...765...16S}. From this process, the molecular dependencies of CNO are extracted.

In order to provide a consistent chemical analysis, we redetermined the chemical abundances assuming as input the effective temperature ($\text{T}_\text{eff}$), surface gravity (log $g$), and metallicity ([Fe/H]) as derived by the ASPCAP pipeline \citep{garciaperez2016}. However, we also applied a simple approach of fixing $\text{T}_\text{eff}$ and log $g$ to values determined independently of spectroscopy, in order to check for any significant deviation in the chemical abundances and to minimize a number of caveats present in ASPCAP/APOGEE-2 abundances for GCs \citep[see, e.g.,][]{2019A&A...622A.191M,2020MNRAS.492.1641M,2021A&A...652A.158R}, as it is affected by a systematic effect that most likely overestimates the $\text{T}_\text{eff}$ values of 2P stars \citep{2021A&A...652A.157G}.

In order to obtain $\text{T}_\text{eff}$ and log $g$ from photometry, we first derived the differential-reddening-corrected CMD of Fig.~\ref{starsdiagram}. We then horizontally projected the position of each observed star until it intersected the isochrone and assumed $\text{T}_\text{eff}$ and log $g$ to be the temperature and gravity of the point on the isochrone that has the same K$_s$ magnitude as the star. 

We underline the fact that for highly reddened objects such as NGC 6569, the interstellar absorption correction depends on the spectral energy distribution of the star, that is, on its temperature. For this reason, we applied a temperature-dependent absorption correction to the isochrone. Without this, it is not possible to obtain a proper fit to the RGB, especially its upper, cooler part. 

Finally, with $\text{T}_\text{eff}$ and log $g$ fixed, we employed the relation from \citet[Eq. 1]{2020A&A...638A..58M} for FGK stars to determine the microturbulence parameter $\xi_t$. The adopted atmospheric parameters are listed in Table~\ref{parameters}.

\begin{table*}[!ht] 
    \caption{Adopted atmospheric parameters inferred from photometry and spectroscopy (uncalibrated and calibrated).}
    \centering
    \scalebox{0.9}{
    \begin{tabular}{l|l|l|l|l|l|l|l|l|l}
    \hline
         ~ & \multicolumn{3}{c|}{Photometry} & \multicolumn{3}{c|}{Uncalibrated} & \multicolumn{3}{c}{Calibrated} \\ \hline \hline
APOGEE ID & T$_\text{eff}$[K] & logg[c.g.s] & $\xi_t$ [km\,s$^{-1}$] & T$_\text{eff}$[K] & logg[c.g.s] & $\xi_t$ [km\,s$^{-1}$] & T$_\text{eff}$[K] & logg[c.g.s] & $\xi_t$ [km\,s$^{-1}$] \\ \hline \hline
2M18132448-3149140 & 3570.0 & 0.850 & 1.772 & 3430.3 & 0.725 & 1.912 & 3597.7 & 0.855 & 1.818 \\ 
2M18133620-3149040 & 4185.9 & 1.139 & 1.579 & 4082.7 & 0.969 & 1.612 & 4247.9 & 0.959 & 1.664 \\ 
2M18133324-3150194 & 4089.8 & 0.967 & 1.611 & 4836.7 & 1.647 & 1.469 & 4954.2 & 1.687 & 1.648 \\ 
2M18134025-3149477 & 3885.1 & 0.592 & 1.692 & 3955.2 & 1.027 & 1.393 & 4117.7 & 1.112 & 1.593 \\ 
2M18133789-3147295 & 4076.6 & 0.942 & 1.619 & 3851.1 & 0.345 & 1.558 & 4010.7 & 0.303 & 1.811 \\ 
2M18134151-3148556 & 4036.5 & 0.869 & 1.628 & 4525.5 & 1.909 & 1.400 & 4670.6 & 1.852 & 1.528 \\ 
2M18133940-3150132 & 3679.8 & 0.213 & 1.780 & 3831.5 & 0.698 & 1.931 & 3994.6 & 0.859 & 2.153 \\ 
2M18133083-3148103 & 4025.2 & 0.848 & 1.636 & 4020.0 & 0.930 & 1.558 & 4019.8 & 0.850 & 1.632 \\ 
2M18133329-3146211 & 4526.9 & 1.754 & 1.471 & 4126.0 & 1.071 & 1.575 & 4126.0 & 1.071 & 1.576 \\ 
2M18134725-3147570 & 4424.6 & 1.568 & 1.500 & 4259.9 & 1.521 & 1.328 & 4442.4 & 1.300 & 1.731 \\ 
2M18135154-3151406 & 4427.3 & 1.572 & 1.496 & 4247.9 & 1.300 & 1.549 & 4419.8 & 1.580 & 1.487 \\ \hline \hline
    \end{tabular}
    }
    \label{parameters}
\end{table*}

\begin{figure}[h!]
    \centering
    \includegraphics[width = 0.5\textwidth, height = 0.6\textheight]{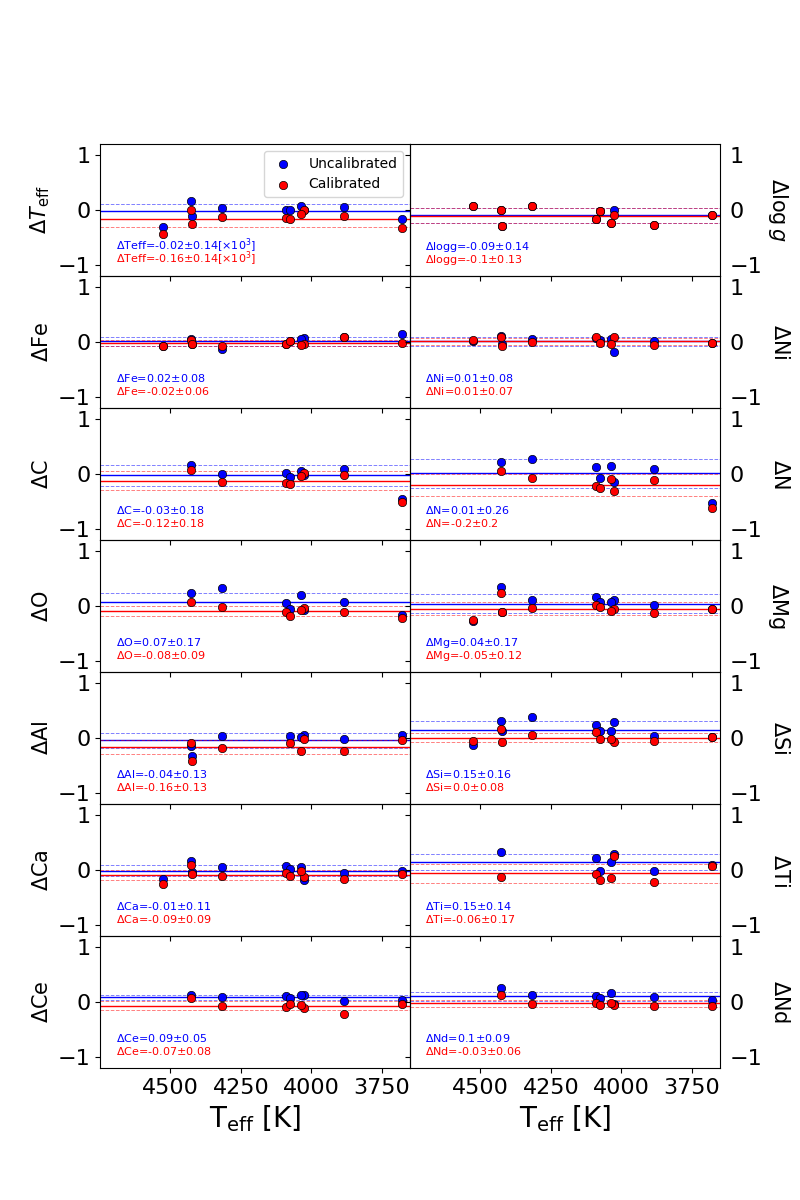}
    \caption{Differences in atmospheric parameters and elemental abundances are shown for three runs adopting photometric and spectroscopic parameters, as listed in Table~\ref{parameters}. The vertical axis refers to the $\Delta$ of each atmospheric parameter and chemical species ([X/Fe]) analyzed in this work. The horizontal axis refers to the photometric $\text{T}_\text{eff}$. The red and blue circles represent the $\Delta$ between photometric-calibrated and photometric-uncalibrated parameters, respectively. The average and standard deviation around the mean of the differences are listed in each panel.
}
    \label{comparison}
\end{figure}

In Figure~\ref{comparison} we compare the abundance sensitivity to the atmospheric parameters derived from photometry with those from the calibrated and uncalibrated ASPCAP parameters. The most significant differences arise in effective temperature, with systematically higher values in the calibrated parameters ($\Delta T_{\text{eff}} = -0.16 \pm 0.14 \times 10^{3}$). 

While iron-peak elements such as Fe ($\Delta \text{Fe} = 0.02 \pm 0.08$ uncalibrated, $-0.02 \pm 0.06$ calibrated) and Ni ($\Delta \text{Ni} = 0.01 \pm 0.08$ uncalibrated, $0.01 \pm 0.07$ calibrated) show relative stability, the largest variations are generally associated with the use of calibrated parameters. Elements such as C ($\Delta \text{C} = -0.12 \pm 0.18$), N ($\Delta \text{N} = -0.2 \pm 0.2$), and Al ($\Delta \text{Al} = -0.16 \pm 0.13$) exhibit significantly larger deviations when using the calibrated values. 

Notably, Si ($\Delta \text{Si} = 0.15 \pm 0.16$) and Ti ($\Delta \text{Ti} = 0.15 \pm 0.14$) also show strong sensitivity, although in these cases, the uncalibrated parameters introduce the larger offsets. These findings indicate that the overestimation of $T_{\text{eff}}$ in the calibrated values leads to an overestimation of the abundances derived by ASPCAP. However, discrepancies with uncalibrated parameters do not have a large impact on our conclusions.

As a consequence, we adopt in the following analysis the atmospheric parameters derived from photometry in order to avoid the known caveats of ASPCAP and to ensure an independent determination of the chemical abundances \citep{2021A&A...652A.157G, 2021A&A...652A.158R}.

\section{Elemental abundances}\label{elementalabundances}
 After we inspected all our spectra visually, we provide reliable abundance determinations for 12 selected chemical species, including those from the iron-peak (Fe, Ni), odd-Z (Al), light elements (C, N), $\alpha$-elements (O, Mg, Si, Ca, Ti), and s-process elements (Ce, Nd). We excluded the remaining chemical species available in APOGEE-2 spectra because their lines are generally very weak and heavily blended with telluric features in the H band. This prevents reliable abundance determinations. 

For this reason, we placed greater emphasis on the elemental abundances of Al, Mg, C, N, and O, which are chemical signatures that are typically used to distinguish stars with different compositions in the MP phenomenon.

The final values of each element that shows reliable lines for each adopted set of atmospheric parameters (photometric and spectroscopic (calibrated and uncalibrated)) are summarized in Table~\ref{abundacnes}. We show the central value for each element with good lines for each atmospheric parameter set: photometry, and uncalibrated and calibrated values from APOGEE DR17. We also compared them with the values estimated by ASPCAP and the mean values determined by \citet{2018AJ....155...71J} and \citet{2011MNRAS.414.2690V}, which are the only previous high-resolution studies of NGC 6569.

We adopted the solar reference abundances from \citet{2005ASPC..336...25A} for most elements, given their wide use and consistency in the literature. However, for the neutron-capture elements Ce and Nd, we used the more recent and detailed determinations provided by \citet{2015A&A...573A..27G}, which include updated atomic data, 3D hydrodynamical model atmospheres, and a rigorous line-by-line analysis. This choice was motivated by the lower systematic uncertainties and better accuracy associated with these heavy elements.

The uncertainties were calculated as the root sum square of the individual uncertainties from errors in each independent atmospheric parameter, 

\begin{equation}\label{error}
\sigma^2 = \sigma^2_{\text{mean}} + \sigma^2_{[\text{X}/\text{H}],\xi_t} + \sigma^2_{[\text{X}/\text{H}],\log g} + \sigma^2_{[\text{X}/\text{H}],\text{T}_{\text{eff}}}.
\end{equation}
Each deviation $\sigma$ represents the change in our initial central value estimation after running BACCHUS with atmospheric parameters varied by $\text{T}_{\text{eff}} \pm 100$ K, $\log g \pm 0.3$ m\,s$^{-2}$, and $\xi_{\text{micro}} \pm 0.05$ km\,s$^{-1}$. This implies running BACCHUS for each of the varied atmospheric parameters and statistically measuring the resulting abundance variations.

\begin{table*}[!ht]
    
    \caption{Elemental abundances of NGC 6569 members.}
    \centering
    \scalebox{0.75}{
    \begin{tabular}{l|l|l|l|l|l|l|l|l|l|l|l|l|l}
    \hline
        Apogee ID & S/N & [C/Fe] & [N/Fe] & [O/Fe] & [Mg/Fe] & [Al/Fe] & [Si/Fe] & [Ca/Fe] & [Ti/Fe] & [Fe/H] & [Ni/Fe] & [Ce/Fe] & [Nd/Fe] \\ \hline \hline
2M18132448-3149140 & 276 & $-0.05$ & $+0.51$ & $+0.34$ & $+0.44$ & $+0.28$ & $+0.20$ & $+0.22$ & $+0.39$ & $-0.83$ & $+0.04$ & $+0.26$ & $+0.41$ \\ 
2M18133940-3150132 & 222 & $-0.56$ & $+0.99$ & $+0.16$ & $+0.46$ & $+0.78$ & $+0.32$ & $+0.33$ & $+0.56$ & $-0.87$ & $+0.07$ & $+0.43$ & $+0.51$ \\ 
2M18134025-3149477 & 174 & $-0.11$ & $+0.34$ & $+0.32$ & $+0.34$ & $+0.21$ & $+0.39$ & $+0.15$ & $+0.13$ & $-0.96$ & $-0.01$ & $+0.19$ & $+0.48$ \\ 
2M18133789-3147295 & 171 & $-0.48$ & $+1.09$ & $+0.30$ & $+0.54$ & $+0.41$ & $+0.38$ & $+0.17$ & $+0.13$ & $-0.93$ & $+0.07$ & $+0.35$ & $+0.49$ \\ 
2M18133324-3150194 & 153 & $-0.29$ & $+1.24$ & $+0.38$ & $+0.49$ & $+0.49$ & $+0.36$ & $+0.35$ & $+0.43$ & $-0.92$ & $+0.10$ & $+0.32$ & $+0.46$ \\ 
2M18133083-3148103 & 130 & $-0.26$ & $+0.90$ & $+0.26$ & $+0.39$ & $+0.40$ & $+0.42$ & $+0.10$ & $+0.55$ & $-0.90$ & $-0.10$ & $+0.34$ & $+0.25$ \\ 
2M18133620-3149040 & 136 & $-0.11$ & $+0.28$ & $+0.39$ & $+0.41$ & $+0.21$ & $+0.28$ & $+0.39$ & $+0.37$ & $-0.83$ & $+0.00$ & $+0.23$ & $+0.47$ \\ 
2M18134151-3148556 & 119 & $-0.12$ & $+0.53$ & $+0.49$ & $+0.34$ & $+0.21$ & $+0.37$ & $+0.38$ & $+0.17$ & $-0.99$ & $+0.04$ & $+0.24$ & $+0.51$ \\ 
2M18134725-3147570 & 64 & -- & -- & -- & $+0.40$ & $+0.37$ & $+0.38$ & $+0.35$ & $+0.62$ & $-0.90$ & $+0.10$ & -- & -- \\ 
2M18135154-3151406 & 60 & $-0.02$ & $+0.41$ & $+0.68$ & $+0.58$ & $+0.35$ & $+0.26$ & $+0.40$ & $+0.37$ & $-0.85$ & $+0.10$ & $+0.18$ & $+0.46$ \\ 
2M18133329-3146211 & 51 & -- & -- & -- & $+0.29$ & $+0.22$ & $+0.31$ & $+0.30$ & -- & $-1.02$ & $-0.04$ & -- & -- \\ \hline
\rowcolor{red!15} Mean & ~ & $-0.22$ & $+0.68$ & $+0.39$ & $+0.43$ & $+0.35$ & $+0.35$ & $+0.28$ & $+0.37$ & $-0.91$ & $+0.03$ & $+0.28$ & $+0.45$ \\ \hline
Std & ~ & $+0.18$ & $+0.34$ & $+0.15$ & $+0.08$ & $+0.17$ & $+0.07$ & $+0.10$ & $+0.18$ & $+0.06$ & $+0.06$ & $+0.08$ & $+0.08$ \\ \hline

        ASPCAP DR17 & ~ & ~ & ~ & ~ & ~ & ~ & ~ & ~ & ~ & ~ & ~ & ~ & ~ \\ \hline \hline
\rowcolor{red!15} Mean & ~ & $-0.21$ & $+0.66$ & $+0.30$ & $+0.32$ & $+0.33$ & $+0.29$ & $+0.28$ & $+0.12$ & $-0.96$ & $+0.04$ & $+0.12$ & -- \\ 
Std & ~ & $+0.16$ & $+0.45$ & $+0.07$ & $+0.04$ & $+0.20$ & $+0.06$ & $+0.12$ & $+0.16$ & $+0.07$ & $+0.03$ & $+0.25$ & -- \\ \hline
Spec(Uncalibrated) & ~ & ~ & ~ & ~ & ~ & ~ & ~ & ~ & ~ & ~ & ~ & ~ & ~ \\ \hline \hline
\rowcolor{red!15} Mean & ~ & $-0.23$ & $+0.74$ & $+0.32$ & $+0.39$ & $+0.41$ & $+0.40$ & $+0.29$ & $+0.19$ & $-0.95$ & $+0.03$ & $+0.20$ & $+0.36$ \\ 
Std & ~ & $+0.10$ & $+0.52$ & $+0.07$ & $+0.12$ & $+0.22$ & $+0.13$ & $+0.10$ & $+0.15$ & $+0.08$ & $+0.06$ & $+0.11$ & $+0.08$ \\ \hline
Spec(Calibrated) & ~ & ~ & ~ & ~ & ~ & ~ & ~ & ~ & ~ & ~ & ~ & ~ & ~ \\ \hline \hline
\rowcolor{red!15} Mean & ~ & $-0.13$ & $+0.95$ & $+0.48$ & $+0.48$ & $+0.53$ & $+0.55$ & $+0.37$ & $+0.39$ & $-0.91$ & $+0.03$ & $+0.36$ & $+0.48$ \\ 
Std & ~ & $+0.09$ & $+0.50$ & $+0.11$ & $+0.06$ & $+0.17$ & $+0.11$ & $+0.09$ & $+0.09$ & $+0.06$ & $+0.10$ & $+0.12$ & $+0.10$ \\ \hline
\citet{2018AJ....155...71J} & ~ & ~ & ~ & ~ & ~ & ~ & ~ & ~ & ~ & ~ & ~ & ~ & ~ \\ \hline \hline
\rowcolor{red!15} Mean & ~ & -- & -- & $+0.38$ & $+0.41$ & $+0.52$ & $+0.35$ & $+0.21$ & -- & $-0.87$ & $-0.08$ & -- & -- \\ 
Std & ~ & -- & -- & $+0.31$ & $+0.09$ & $+0.14$ & $+0.09$ & $+0.10$ & -- & $+0.06$ & $+0.05$ & -- & -- \\ \hline
\citet{2011MNRAS.414.2690V} & ~ & ~ & ~ & ~ & ~ & ~ & ~ & ~ & ~ & ~ & ~ & ~ & ~ \\ \hline \hline
\rowcolor{red!15} Mean & ~ & $-0.27$ & -- & $+0.48$ & $+0.50$ & $+0.38$ & $+0.49$ & $+0.31$ & $+0.40$ & $-0.79$ & -- & -- & -- \\ 
Std & ~ & $+0.11$ & -- & $+0.09$ & $+0.06$ & $+0.11$ & $+0.08$ & $+0.04$ & $+0.04$ & $+0.05$ & -- & -- & -- \\ \hline \hline

    \end{tabular}
    }
    \tablefoot{
    For each species we show the mean value and the standard deviation. We compare with values estimated with ASPCAP/APOGEE DR17 and with previous high-resolution studies on this cluster \citet{2018AJ....155...71J} and \citet{2011MNRAS.414.2690V}. The Solar reference abundances are from \citet{2005ASPC..336...25A}, except for Ce and Nd, which is taken from \citet{2015A&A...573A..27G}.}
    \label{abundacnes}
\end{table*}

\subsection{Iron-peak elements Fe and Ni}
We measure a mean metallicity of $\langle \mathrm{[Fe/H]} \rangle = -0.91 \pm 0.06\ (1\sigma) \pm 0.018\ (\mathrm{std}/\sqrt{N})$, with a dispersion of $0.06 \pm 0.017\ (\mathrm{std}/\sqrt{2 \times N})$ dex. 

\begin{figure*}[h!]
    \centering
    \includegraphics[width = 1\textwidth, height = 0.35\textheight]{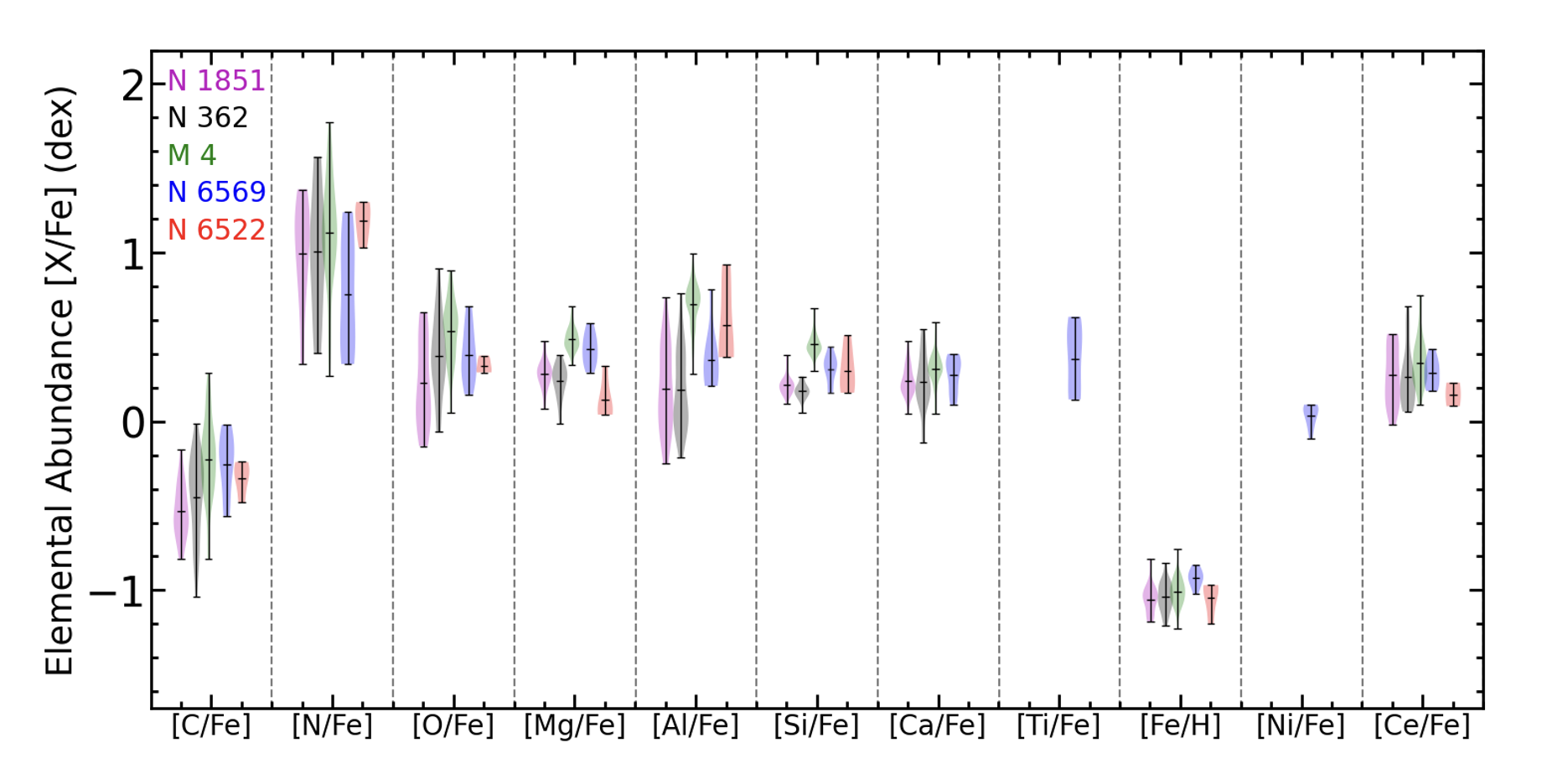}
    \caption{[X/Fe] and [Fe/H] abundance density estimation (violin representation) of NGC 6569 (blue), compared to
elemental abundances of Galactic GCs NGC 1851 (magenta), NGC 362 (black), and M4 (green) from \citet{2020MNRAS.492.1641M} and the bulge GC NGC 6522 (red) from \citet{2019A&A...627A.178F}.}
    \label{violin}
\end{figure*}

Table~\ref{abundacnes} lists a relatively large total [Fe/H] range (0.19 dex) for NGC 6569, which is mainly driven by the low metallicity ([Fe/H] = $-1.02$) of the star 2M18133329-3146211, whose deviation from the mean is slightly larger than the typical internal errors ($\sigma_{\mathrm{[Fe/H]}} \sim 0.04$–$0.08$ dex). Moreover, the [Fe/H] abundance ratios listed in Table~\ref{abundacnes} show that the observed dispersion (0.06 dex) agrees well with the measurement errors (< 0.08 dex), so we find no evidence for a statistically significant metallicity spread—this is a common behavior in GCs (both BGCs and non-BGCs) with similar metallicities (see Fig.~\ref{violin}).

This fact underlines that we have avoided the caveat introduced by ASPCAP, where 1P and 2P stars do not share the same metallicity (see Fig. 5 of \citet{2021A&A...652A.157G}, Geisler et al., in prep). Our [Fe/H] is slightly more metal-poor than the $\mathrm{[Fe/H]} = -0.87$ estimated by \citet{2018AJ....155...71J}, and significantly more metal-poor than the value obtained by \citet{2011MNRAS.414.2690V}, $\mathrm{[Fe/H]} = -0.79$. 

Our [Fe/H] is also significantly more metal-poor than the value tabulated by \citet{2010arXiv1012.3224H} ([Fe/H] = –0.76) and the estimate by \citet{2005MNRAS.361..272V} ([Fe/H] = –0.88), which linked metal abundance to a variety of NIR indices measured along the RGB. However, it is slightly more metal-rich than the ASPCAP estimate ([Fe/H] = –0.96). 

As our [Fe/H] determination was estimated directly from Fe I atomic lines and high-resolution spectra, it is likely more precise than the literature estimations. The use of Fe I atomic lines ensures that the measurement is directly related to the iron content, minimizing the uncertainties associated with indirect methods, as these lines are typically strong and less affected by non-local thermodynamic equilibrium (NLTE) effects compared to other lines \citep{Ssmith2021}.

Additionally, we found a $\langle\mathrm{[Ni/Fe]}\rangle$ ratio slightly higher than the solar value. This is the second estimation of [Ni/Fe] for this cluster, and overall there is good agreement with the measurement reported by \citet{2018AJ....155...71J}, who obtained an average of $\langle\mathrm{[Ni/Fe]}\rangle = 0.03$. Our measurements yield an average $\langle\mathrm{[Ni/Fe]}\rangle = 0.03 \pm 0.06\ (1\sigma) \pm 0.018\ (\mathrm{std}/\sqrt{N})$, with a dispersion of $0.06 \pm 0.017\ (\mathrm{std}/\sqrt{2 \times N})$ dex. 

From Table~\ref{abundacnes}, we infer no intrinsic spread in Ni, as the observed dispersion (0.06 dex) agrees well with the measurement errors (0.06–0.15 dex). This suggests that the variability in [Ni/Fe] can be largely attributed to measurement errors rather than intrinsic differences in the nickel content of the stars, similar to what is observed in GCs with comparable metallicities \citep{gratton2004abundance}.

\subsection{Light elements C and N}
NGC 6569 exhibits a strong enrichment in N, with a mean $\langle \mathrm{[N/Fe]} \rangle = +0.68 \pm 0.34\ (1\sigma) \pm 0.10\ (\mathrm{std}/\sqrt{N})$, and a large star-to-star spread of $+0.96$ dex. Almost all the stars examined in NGC 6569 are enriched in N and depleted in C with respect to the solar value, with a mean $\langle \mathrm{[C/Fe]} \rangle = -0.22 \pm 0.18\ (1\sigma) \pm 0.05\ (\mathrm{std}/\sqrt{N})$, and a large star-to-star spread of $+0.56$ dex.

Figure~\ref{abundiag}(a) and (c) reveal a statistically significant C–N anticorrelation and N–Al correlation. The anticorrelation between C and N, with a Pearson correlation coefficient of $r_{C\text{--}N} = -0.80$, indicates a strong inverse relation between these two elements. The corresponding p value of $p = 0.005$ is below the typical significance threshold of $\sim$ 0.05, confirming that the anticorrelation is statistically significant. This means that as the N abundance increases, the C abundance decreases significantly among the stars in NGC 6569. 

The first dredge-up alters the surface abundances of N and C by bringing processed material to the surface during the red giant phase. This decreases the C/N ratio and affects the $^{12}\text{C}/^{13}\text{C}$ ratios. However, it does not cause the observed C–N anticorrelation in globular clusters such as NGC 6569, which is instead attributed to early enrichment processes involving material from previous generations of massive stars \citep{2024MNRAS.530..149R}, as all of our stars are well above the RGB bump, where the first dredge-up occurs.

Similarly, the correlation between N and Al, with a Pearson correlation coefficient of $r_{N\text{--}Al} = 0.73$, indicates a moderate positive relation between these elements. The p value for this correlation is $ \sim p = 0.02$, which is below the 0.05 threshold, suggesting that the correlation is statistically significant. This implies that as the N abundance increases, the Al abundance tends to increase as well, as is generally seen in other GCs.

The chemical trends of these elements are similar to those observed in Galactic GCs and BGCs, such as NGC 6522, as shown in Fig.~\ref{violin}, but with a smaller spread, indicating similar stellar processes at play. NGC 6569 contains at least two groups of stars that likely correspond to a first stellar population characterized by $\mathrm{[N/Fe]} \lesssim +0.7$ and a second population with $\mathrm{[N/Fe]} \gtrsim +0.7$. 

The MPs in NGC 6569 are characterized by their N abundances for the first time, and C and N are also measured simultaneously for the first time. This allowed us to show the N–C anticorrelation clearly. This anticorrelation is a typical feature of stars in GCs and provides clear evidence of multiple stellar populations, one of which is N depleted and C enriched (1P stars), and the other is N enriched and C depleted (2P stars) \citep[see, e.g.,][]{2020MNRAS.492.1641M}.

\subsection{$\alpha$-elements, O, Mg, Ca, Si, and Ti}

Figure~\ref{violin} and Table~\ref{abundacnes} show that NGC 6569 exhibits significant $\alpha$-element enhancement, with mean values ranging from +0.28 to +0.43 for [O/Fe], [Mg/Fe], [Si/Fe], [Ca/Fe], and [Ti/Fe]. The mean values for these elements are as follows: $\langle \mathrm{[O/Fe]} \rangle = +0.39 \pm 0.15$, $\langle \mathrm{[Mg/Fe]} \rangle = +0.43 \pm 0.08$, $\langle \mathrm{[Si/Fe]} \rangle = +0.35 \pm 0.07$, $\langle \mathrm{[Ca/Fe]} \rangle = +0.28 \pm 0.10$, and $\langle \mathrm{[Ti/Fe]} \rangle = +0.37 \pm 0.18$. 

This agrees well with GCs of similar metallicity (see Fig.~\ref{violin}) \citep{2020MNRAS.492.1641M}. The $\alpha$-elements display minimal star-to-star variation within the typical uncertainties. For example, the observed dispersions for these elements are 0.15 dex for [O/Fe], 0.08 dex for [Mg/Fe], 0.07 dex for [Si/Fe], 0.10 dex for [Ca/Fe], and 0.18 dex for [Ti/Fe], which are comparable to or smaller than the measurement errors (0.16 dex for [O/Fe], 0.17 dex for [Mg/Fe], 0.18 dex for [Si/Fe], 0.11 dex for [Ca/Fe], and 0.13 dex for [Ti/Fe]). 

This indicates that the intrinsic scatter in the abundances of these elements is small and consistent with the measurement uncertainties. The homogeneity in the $\alpha$-element abundances within the cluster suggests that star formation occurred rapidly, before the gas was significantly mixed with the surrounding interstellar medium. The lack of significant variation in the $\alpha$-element abundances implies a well-mixed environment in which the enrichment processes were uniform. This supports the idea of a single, rapid burst of star formation followed by the pollution of the interstellar medium by type II supernovae \citep{Bastian_2018}.

Notably, we find a [Si/Fe] value that closely matches that of \citet{2018AJ....155...71J}, who reported a mean of [Si/Fe] = 0.35 in NGC 6569, but it is lower than the value found by \citet{2011MNRAS.414.2690V}, who measured [Si/Fe] = 0.49.

As expected for metal-rich GCs such as NGC 6569, we did not observe a clear Mg–Si anticorrelation, which is typical of higher-metallicity clusters and indicates no observable 2P star enrichment features \citep[e.g.,][]{2019A&A...622A.191M}. According to \citet{2020MNRAS.492.1641M}, this is likely due to the lower efficiency of Al–Si production from the Mg–Al cycle in this metallicity regime: In lower-metallicity stars, the Mg–Al cycle can operate at higher temperatures, allowing Si production through high-temperature proton-capture reactions (e.g., $^{26}$Al(p,$\gamma$)$^{27}$Si(e$^{-}$,$\nu$)$^{27}$Al(p,$\gamma$)$^{28}$Si).

The $\alpha$-elements in NGC 6569 are overabundant compared to the Sun ([$\alpha$/Fe]). This is a common trait among Galactic GCs and field stars with similar metallicities and shows that NGC 6569 is consistent with other Galactic GCs of similar metallicity in terms of $\alpha$-element content \citep{10.1007/s00159-012-0050-3}. BGCs follow similar trends (see Fig.~\ref{violin}). The high abundance of $\alpha$-elements compared to the Sun in NGC 6569 suggests an early enrichment by type II supernovae, which produce these elements.

\subsection{The odd-Z element Al}
\begin{figure*}[h!]
    \centering
    \includegraphics[width=\textwidth]{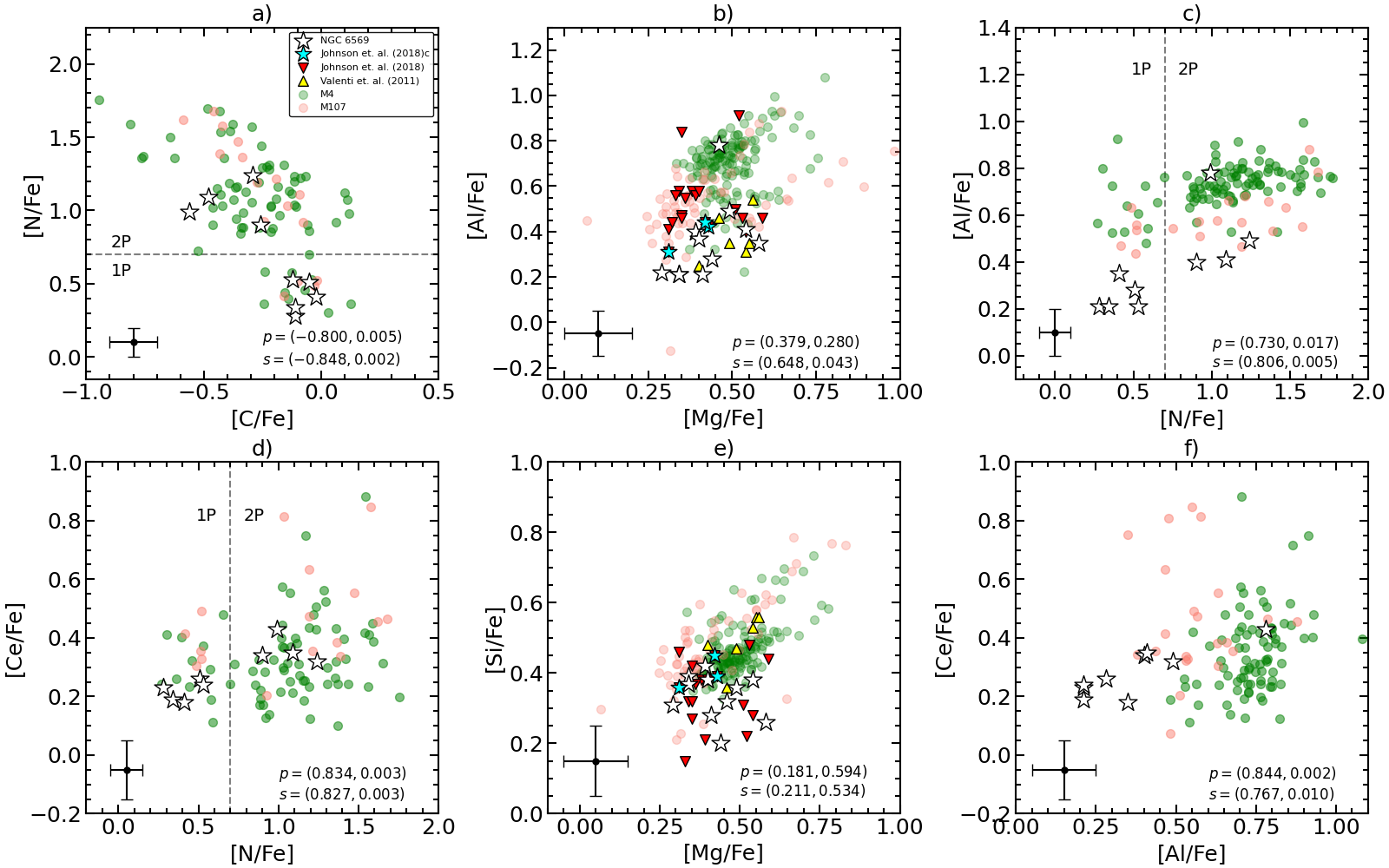}
    \caption{Panels (a)–(f): [C/Fe]–[N/Fe], [Mg/Fe]–[Al/Fe], [Al/Fe]–[N/Fe], [N/Fe]–[Ce/Fe], [Si/Fe]–[Mg/Fe], and [Al/Fe]–[N/Fe] distributions for NGC 6569 including this work (white stars), previous high-resolution studies (\citet{2011MNRAS.414.2690V} (yellow triangles) and \citet{2018AJ....155...71J} (red triangles), and M4 (green circles) and M107 (orange circles) from \citet{2020MNRAS.492.1641M}. We also show as cyan stars the three stars in common with \citet{2018AJ....155...71J}. The dashed black line in panels a), c), and d) represents the division between 1P and 2P at [N/Fe] = 0.7. The typical uncertainties for NGC 6569 members are also shown, as described in Section \ref{elementalabundances}. Additionally, we display the Pearson and Spearman correlation coefficients along with their corresponding p values in each panel.}
    \label{abundiag}
\end{figure*}

NGC 6569 exhibits a mean Al enrichment of $\langle \mathrm{[Al/Fe]} \rangle = +0.35 \pm 0.17$, with a dispersion of $0.17 \pm 0.05$ dex. We observed no significant variation in $\mathrm{[Al/Fe]}$ beyond typical errors (0.08 dex). One star showed extreme enrichment at [Al/Fe] = 0.78, which mainly drives the observed dispersion. This enrichment correlates with 2P stars based on their N content, as indicated by the Pearson correlation coefficient for Al–N of $r_{\mathrm{Al\text{--}N}} = 0.73$ with a p value of $p = 0.017$, indicating a moderately significant statistical correlation. The observed dispersion prevents us from classifying MPs based on Al in NGC 6569, however, as is commonly done \citep[e.g.,][]{2020MNRAS.492.1641M}. Given the small sample size, we were unable to identify any clear $\mathrm{[Al/Fe]}$ variation within NGC 6569. In contrast, the higher values of $\mathrm{[Al/Fe]} = 0.52$ found by \citet{2018AJ....155...71J}, similar to those derived from calibrated atmospheric parameters ([Al/Fe] = 0.52; see Fig.~\ref{abundiag}), may result from their larger sample of 18 stars and/or a higher fraction of 2P stars. We suggest that this discrepancy is likely due to a combination of both factors: The larger sample and the higher effective temperatures inferred by \citet{2018AJ....155...71J}.

Figure~\ref{abundiag} reveals a statistically significant Al–Ce correlation, with a coefficient of $r_{\mathrm{Al\text{--}Ce}} = 0.84$ and a p value of $p = 0.002$, suggesting a potential link between Al and Ce abundances that may indicate enrichment by similar processes \citep{2022A&A...658A.116F}.

The comparison NGC 6569 with other clusters of similar metallicity (see Fig.~\ref{violin}) reveals that the Al values in NGC 6569 are less widely dispersed, but have a similar central value. This lower dispersion in [Al/Fe] in NGC 6569 might be caused by several factors. One possible explanation is a more uniform initial mass function (IMF) and star formation history in NGC 6569 than in NGC 1851 and NGC 362. A more uniform IMF would result in more homogeneous enrichment of Al from intermediate-mass stars during their AGB phase, leading to a weaker variation in Al content among the stars \citep{2015MNRAS.447.1033M}.

Another factor might be the efficiency of mixing processes within the cluster. If NGC 6569 experienced more efficient mixing of its interstellar medium, it would result in a more homogeneous distribution of elements such as Al \citep{Sha2011Nanostructure}. Additionally, the specific evolutionary paths and stellar interactions within NGC 6569 could differ from those in NGC 1851 and NGC 362, which would contribute to the observed differences in Al dispersion \citep{Navin2015New}. 

This is supported by the more uniform enrichment seen in BGCs such as NGC 6522 (see Fig.~\ref{violin}), suggesting a different evolutionary history compared with non-BGCs. Moreover, the small spread could also be due a relatively small fraction of 2P stars. 
The Mg–Al anticorrelation is notably absent in high-metallicity clusters such as NGC 6569 (see Fig.~\ref{abundacnes}). This absence is likely explained by the efficiency of the Mg–Al cycle, which operates at the higher temperatures found in lower-metallicity environments. In high-metallicity clusters, the temperatures in stellar interiors are not sufficient to activate the Mg–Al cycle to the same extent, leading to a lack of significant Mg depletion and Al enhancement. As a result, no Mg–Al anticorrelation is observed in these clusters \citep{pancino2017gaia}. 

This phenomenon supports the idea that the chemical enrichment processes and formation history of globular clusters are heavily influenced by their metallicity, with higher-metallicity clusters showing different nucleosynthesis pathways compared to their lower-metallicity counterparts \citep[e.g.,][]{2020MNRAS.492.1641M}.

\subsection{n-capture elements Ce and Nd}
We find a mean $\langle \mathrm{[Ce/Fe]} \rangle = +0.28 \pm 0.08$, which is highly overabundant compared to the Sun, but comparable to the Ce levels observed in other Galactic GCs and BGCs with a similar metallicity \citep[e.g.,][]{2019A&A...622A.191M, 2020MNRAS.492.1641M}, as also shown in Fig.~\ref{violin}. The observed spread in the s-process element Ce (0.08 dex) does not exceed the observational uncertainties (0.08–0.18 dex).

In this context, Figure~\ref{abundiag} suggests a possible N–Ce correlation, with a Pearson correlation coefficient of $r_{\mathrm{N\text{--}Ce}} = 0.83$ and a p value of $p = 0.003$, indicating a statistically significant correlation. This feature has previously been noted in only a few high-metallicity BGCs, such as NGC 6380 \citep{2021ApJ...918L...9F} and Ton 2 \citep{2022A&A...658A.116F}. 

The significant correlations among Al, N, and Ce support a scenario in which these elements are enriched through similar astrophysical processes, possibly involving contributions from asymptotic giant branch (AGB) stars. The possible N–Ce correlation can be interpreted in this context: During the AGB phase, stars undergo thermal pulses that produce and mix s-process elements such as Ce. These stars can also produce N through hot-bottom burning processes. Thus, a correlation between N and Ce in a GC would support the scenario in which AGB stars significantly contribute to the chemical enrichment of the cluster \citep{2022A&A...658A.116F}.

We also find a mean $\langle \mathrm{[Nd/Fe]} \rangle = +0.45 \pm 0.08$, which is highly overabundant compared to the Sun, but similar to the levels of r-process elements (such as Eu) observed in other Galactic GCs with a comparable metallicity \citep[e.g.,][]{1996AJ....112.1517S, 2018AJ....155...71J}. Although this is not a direct comparison, the similar trends of r-process elements in high-metallicity GCs may indicate a common enrichment history \citep{bekki2017formation}.

\begin{figure}[h!]
    \centering
    \includegraphics[scale=0.45]{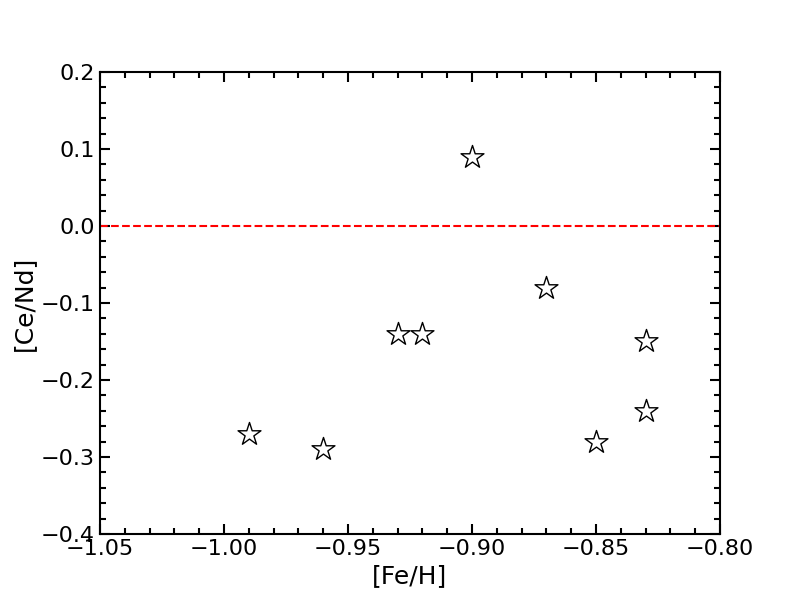}
    \caption{[Ce/Nd] ratio against [Fe/H] for NGC 6569 members with reliable measurements.}
    \label{ratio}
\end{figure}

Nd is reported in NGC 6569 for the first time here. Following an approach similar to that by \citet{2018AJ....155...71J}, who classified the possible sources of n-capture elements in NGC 6569 based on their elemental ratios, we found a negative [Ce/Nd] ratio of [Ce/Nd] = $-0.17 \pm 0.12$. Given the different contributions from neutron-capture processes to Ce and Nd (e.g., in the Solar System, Ce and Nd are s-process-dominated elements; their s-/r-process contributions to the pre-solar composition are estimated as 85/15\% for Ce and 62/38\% for Nd \citep{Prantzos_2019}), the abundance ratio [Ce/Nd] in stellar populations can indicate the dominant production process.

A high [Ce/Nd] ratio signifies a greater proportion of s-process contribution, while a low [Ce/Nd] ratio suggests formation dominated by the r-process. \citet{2018AJ....155...71J} measured La and Eu, which are associated with the s- and r-processes, respectively, and found a negative average [La/Eu] ratio of $-0.11$. Figure~\ref{ratio} shows a similar behavior in NGC 6569. One star appears to be significantly affected by AGB star polluters (i.e., [Ce/Nd] $>$ 0), while the rest indicate substantial enrichment by r-process material.

Although n-capture elements are not exclusively produced by a single process but rather by a combination of sources \citep{2024ApJ...965..119S}, the [Ce/Nd] ratio in NGC 6569 clearly suggests a predominance of high-energy process origins. The presence of higher Nd relative to Ce aligns with a scenario in which r-process contributions are significant, possibly indicating enrichment from neutron star mergers or other r-process sites.

\section{Concluding remarks}\label{conclusion}

We have performed a high-resolution (R$\sim $22,500) spectral analysis in the H band of 11 giant star members of the BGC NGC 6569 as part of the CAPOS survey using APOGEE-2 data. We examined 12 chemical species belonging to the light (C and N), $\alpha$ (O, Mg, Si, Ca, and Ti), iron-peak (Fe and Ni), odd-Z (Al), and s-process (Ce and Nd) elements with the code BACCHUS. In addition, we derived a mean radial velocity of RV $ = -49.75 \pm 3.68$ km s$^{-1}$, which is consistent with previous measurements by \citet{2018AJ....155...71J} (RV $ = -48.8 \pm 5.3$ km s$^{-1}$) and \citet{2021MNRAS.505.5957B} and indicates a robust membership for our sample stars. The cluster was found to have a distance of $\text{d}_\odot$ = 12.4 $\pm$ 1.45 kpc, an extinction $A_V = 1.5$, and a reddening value of $E(B-V) = 0.682$, which is slightly higher than reported previously. Our findings are summarized below.

\begin{itemize}
    \item We presented the first detailed characterization of MPs based on [N/Fe] and [C/Fe] abundances in NGC 6569 that were simultaneously derived from molecular features. It clearly demonstrates the presence of both 1P and 2P stars.
    
    \item NGC 6569 exhibits strong enrichment in N, with a mean $\langle$[N/Fe]$\rangle = +0.68 \pm 0.34$ and a spread of 0.90 dex, and corresponding C depletion with $\langle$[C/Fe]$\rangle = -0.22 \pm 0.18$ and a spread of 0.45 dex, clearly indicating the presence of multiple stellar populations within the cluster.
    
    \item Our results showed a statistically significant anticorrelation between [C/Fe] and [N/Fe], with a Pearson coefficient $r = -0.80$ and $p = 0.02$. This confirms the typical C–N anticorrelation observed in globular clusters \citep[e.g., ][]{2020MNRAS.492.1641M}.
    
    \item NGC 6569 shows enhancement in $\alpha$-elements, with a combined mean [$\alpha$/Fe] = $+0.36 \pm 0.06$, consistent with other globular clusters of similar metallicity \citep[][]{2020MNRAS.492.1641M}.
    
    \item The low star-to-star variation in $\alpha$-elements (with dispersions between 0.08 and 0.18 dex) suggests a rapid and homogeneous star formation that was likely driven by early enrichment from type II supernovae.
    
    \item The absence of the Mg–Al anticorrelation in this high-metallicity cluster supports the notion that the Mg–Al cycle is less efficient at higher metallicities. This is consistent with observations in other high-metallicity clusters \citep[e.g., ][]{pancino2017gaia}.
    
    \item NGC 6569 exhibits overabundances of Ce and Nd, with $\langle$[Ce/Fe]$\rangle = +0.28 \pm 0.08$ and $\langle$[Nd/Fe]$\rangle = +0.45 \pm 0.08$, which is consistent with the levels observed in other Galactic GCs with a similar metallicity \citep[e.g., ][]{2019A&A...622A.191M}.
    
    \item The observed [Ce/Nd] ratio of $\langle$[Ce/Nd]$\rangle = -0.17 \pm 0.12$ implies a predominance of r-process material in the cluster, which might indicate contributions from neutron star mergers or other high-energy processes \citep[e.g., ][]{bekki2017formation}.
    
    \item A potential [N/Fe]–[Ce/Fe] correlation suggests a possible link between N and Ce abundances, which could indicate contributions from AGB stars during the cluster formation.
    
    \item The mean metallicity of NGC 6569 is determined to be $\langle$[Fe/H]$\rangle = -0.91 \pm 0.06$, and no statistically significant metallicity spread is observed. This indicates a homogeneous Fe content among the cluster stars.
    
    \item Our [Fe/H] value is slightly more metal poor than was reported by \citet{2018AJ....155...71J} ($\mathrm{[Fe/H]} = -0.87$) and significantly lower than \citet{2011MNRAS.414.2690V} ($\mathrm{[Fe/H]} = -0.79$). However, our measurements are more metal rich than the ASPCAP estimates ($\mathrm{[Fe/H]} = -0.96$), although they were based on the same APOGEE-2 spectra.
    
    \item The [Ni/Fe] ratio is slightly higher than solar, with a mean $\langle$[Ni/Fe]$\rangle = +0.03 \pm 0.06$. This agrees well with \citet{2018AJ....155...71J} and further supports the homogeneity of Ni in the cluster.
    
    \item NGC 6569 exhibits a mean Al enrichment of $\langle$[Al/Fe]$\rangle = +0.35 \pm 0.17$ without significant star-to-star variation. This suggests a homogeneous distribution of Al within the cluster.
    
    \item Our Al abundance is lower than the value found by \citet{2018AJ....155...71J} ($\mathrm{[Al/Fe]} = +0.52$), but it is consistent with the value reported by \citet{2011MNRAS.414.2690V} ($\mathrm{[Al/Fe]} = +0.38$), as well as with the ASPCAP estimates.
\end{itemize}

\begin{acknowledgements}
S.V. gratefully acknowledges the support provided by Fondecyt regular n. 1220264, and by the ANID BASAL projects ACE210002 and FB210003. J.G.F.-T. gratefully acknowledges the grant support provided by Proyecto Fondecyt Iniciación No. 11220340, from ANID Concurso de Fomento a la Vinculación Internacional para Instituciones de Investigación Regionales (Modalidad corta duración) Proyecto No. FOVI210020, and also from the grant support from the Joint Committee ESO-Government of Chile 2021 (ORP 023/2021). D.G. gratefully acknowledges support from the FONDECYT regular n. 1220264. D.G. also acknowledges financial support from the Dirección de Investigación y Desarrollo de la Universidad de La Serena through the Programa de Incentivo a la Investigación de Académicos (PIA-DIDULS). C.M. thanks the support provided by  GEMINI-ANID  Postdoctorado No.32230017. Funding for the Sloan Digital Sky Survey IV has been provided by the Alfred P. Sloan Foundation, the U.S. Department of Energy Office of Science, and the Participating Institutions. SDSS-IV acknowledges support and resources from the Center for High-Performance Computing at the University of Utah. The SDSS web site is www.sdss.org. SDSS-IV is managed by the Astrophysical Research Consortium for the Participating Institutions of the SDSS Collaboration including the Brazilian Participation Group, the Carnegie Institution for Science, Carnegie Mellon University, the Chilean Participation Group, the French Participation Group, Harvard-Smithsonian Center for Astrophysics, Instituto de Astrofísica de Canarias, The Johns Hopkins University, Kavli Institute for the Physics and Mathematics of the Universe (IPMU)/University of Tokyo, Lawrence Berkeley National Laboratory, Leibniz Institut für Astrophysik Potsdam (AIP), Max-Planck-Institut für Astronomie (MPIA Heidelberg), Max-Planck-Institut für Astrophysik (MPA Garching), Max-Planck-Institut für Extraterrestrische Physik (MPE), National Astronomical Observatory of China, New Mexico State University, New York University, University of Notre Dame, Observatório Nacional/MCTI, The Ohio State University, Pennsylvania State University, Shanghai Astronomical Observatory, United Kingdom Participation Group, Universidad Nacional Autónoma de México, University of Arizona, University of Colorado Boulder, University of Oxford, University of Portsmouth, University of Utah, University of Virginia, University of Washington, University of Wisconsin, Vanderbilt University, and Yale University. This work has made use of data from the European Space Agency (ESA) mission Gaia (http://www.cosmos.esa.int/gaia), processed by the Gaia Data Processing and Analysis Consortium (DPAC, http://www.cosmos.esa.int/web/gaia/dpac/consortium). Funding for the DPAC has been provided by national institutions, in particular the institutions participating in the Gaia Multilateral Agreement. This publication also makes use of data products from the Two Micron All Sky Survey (2MASS), a joint project of the University of Massachusetts and the Infrared Processing and Analysis Center/California Institute of Technology, funded by the National Aeronautics and Space Administration and the National Science Foundation (https://irsa.ipac.caltech.edu/Missions/2mass.html). We acknowledge the use of software tools such as \texttt{numpy} \citep{2020Natur.585..357H}, \texttt{matplotlib} \citep{4160265}, and \texttt{TOPCAT} \citep{2005ASPC..347...29T}, which were essential for the data analysis and the generation of figures in this work.
\end{acknowledgements}

\bibliography{bib}

\end{document}